\documentclass{article}

\usepackage{PRIMEarxiv}

\usepackage[utf8]{inputenc} 
\usepackage{amsmath}
\usepackage{tabularx}
\usepackage[nice]{nicefrac}
\usepackage{algpseudocode}
\usepackage{units}
\usepackage[T1]{fontenc}    
\usepackage{hyperref}       
\usepackage{url}            
\usepackage{booktabs}       
\usepackage{amsfonts}       
\usepackage{microtype}      
\usepackage{lipsum}
\usepackage{fancyhdr}       
\usepackage{graphicx}       
\graphicspath{{media/}}     
\usepackage{multirow}
\usepackage{makecell}
\usepackage{algorithm}
\usepackage{titlesec}

\usepackage{authblk}
\title{Dynamic Control of Spontaneous Emission Using Magnetized InSb Higher-Order-Mode Antennas}
\author[1]{Sina Aghili}
\author[2,3]{Rasoul Alaee}
\author[4]{Amirreza Ahmadnejad}
\author[2]{Ehsan Mobini}
\author[4]{Mohamadreza Mohamadpour}
\author[3,5]{Carsten Rockstuhl}
\author[2,6]{Robert W. Boyd}
\author[1,2]{Ksenia Dolgaleva}

\affil[1]{School of Electrical Engineering and Computer Science, University of Ottawa, Ottawa, Ontario, K1N 6N5, Canada}
\affil[2]{Department of Physics, University of Ottawa,
Ottawa, Ontario K1N 6N5, Canada}
\affil[3]{Karlsruhe Institute of Technology, Institute of Theoretical Solid State Physics, D-76131 Karlsruhe, Germany}
\affil[4]{Department of Electrical Engineering, Sharif University of Technology, Tehran 11155-4365, Iran}
\affil[5]{Institute of Nanotechnology, Karlsruhe Institute of Technology, D-76344 Eggenstein-Leopoldshafen,
Germany}
\affil[6]{The Institute of Optics, University of Rochester, Rochester, New York 14627, United
States}

\titleformat{\section}[block]{\normalfont\Large\bfseries}{\thesection\hspace{0.5em}}{0em}{}[]

\begin{document}
\maketitle

\begin{abstract} 
We exploit InSb's magnetic-induced optical properties to propose THz sub-wavelength antenna designs that actively tune the radiative decay rates of dipole emitters at their proximity. The proposed designs include a spherical InSb antenna and a cylindrical Si-InSb hybrid antenna that demonstrate distinct behaviors; the former dramatically enhances both radiative and non-radiative decay rates in the epsilon-near-zero region due to the dominant contribution of the Zeeman splitting electric octupole mode. The latter realizes significant radiative decay rate enhancement via magnetic octupole mode, mitigating the quenching process and accelerating the photon production rate. A deep learning-based optimization of emitter positioning further enhances the quantum efficiency of the proposed hybrid system. These novel mechanisms are potentially promising for tunable THz single-photon sources in integrated quantum networks.
\end{abstract}

\keywords{Active antenna, Indium Antimonide (InSb), Local density of states (LDOS), Multipole moments,
Radiative decay rate, Zeeman splitting effect, III-V semiconductors}
\section{Introduction}
It has been long appreciated that the local density of states (LDOS), as seen by a quantum emitter, can be altered in a structured electromagnetic environment  \cite{drexhage1970influence,purcell1946resonance}. The spontaneous emission (SE) rate and transition dipole strength, as the functional characteristics of a quantum emitter, greatly rely on LDOS variations \cite{glauber1991quantum}.
In simple terms, the electric or magnetic dipole transition of quantum emitters, when considered as two-level systems, can couple to photonic resonances. Then, the enhancement of the LDOS, thanks to the structured photonic environment, accelerates the spontaneous transition rates of the quantum emitter. The SE rate enhancement is of great interest to diverse applications, including quantum source emission engineering \cite{paesani2020near,choi2016engineering,wang2020enhanced}, ultra-brighter optoelectronic devices \cite{arbel2011light,romeira2018purcell,wang2018enhancing,yu2019nanowire}, enhanced light-matter interaction for fluorescence spectroscopy \cite{russell2012large,tam2007plasmonic,cambiasso2017bridging}, and single-photon sources (SPSs) \cite{aharonovich2016solid,buckley2012engineered,lohrmann2017review}. 

Optical resonators are high-potential devices that can serve as the structured electromagnetic media for quantum emitters by providing an enhanced LDOS \cite{purcell1995spontaneous,haroche2014exploring}. Radiative coupling channels emerge in a coupled emitter-resonator system, allowing one to improve the SE rate of the quantum emitter. Microcavities \cite{qian2021spontaneous}, photonic crystals, and hyperbolic metamaterials (HMMs) \cite{inam2018hyperbolic,ahmed2020hyperbolic,guclu2014radiative} are closed-cavity platforms that have been intensively investigated for their potential to increase the SE rate of quantum emitters. In recent decades, optical antennas have been the subject of this extensive research. Optical antennas are open resonators composed of sub-wavelength dielectric or metallic elements that transfer the energy of their resonance modes to quantum emitters via radiative and non-radiative channels  \cite{bonod2020controlling,agio2012optical,taminiau2008optical}.
Enhanced light-matter interaction at the deep sub-wavelength scale is the distinguishing advantage of antennas over cavities, leading to LDOS enhancement of quantum emitters in the coupled emitter-antenna systems. In this direction, metallic antennas enable one to modify the SE rate of quantum emitters due to the high near-field enhancement attained by localized surface plasmon resonances (LSPRs) \cite{kuhn2006enhancement,anger2006enhancement,mi2019control,jiang2021single,aghili2018proposing}. Metallic nanorods \cite{rogobete2007design} and ring resonators \cite{hein2013tailoring} are highly effective plasmonic subwavelength antennas for enhancing the SE rate of quantum emitters through the excitation of LSPRs, including transversal and longitudinal plasmonic modes in nanorods, and concurrent electric and magnetic resonant modes in ring resonators. In general, coupled emitter-plasmonic antenna systems usually introduce dominant non-radiative channels due to high Joule heating losses of metallic building blocks, resulting in low quantum efficiency for the coupled systems.

Alternative designs have been investigated to mitigate dissipation losses of plasmonic antennas. Interestingly, it was shown that plasmonic devices using a dielectric spacer could avoid or limit non-radiative decay channels \cite{dhawan2020extreme}. This idea has flourished by employing dielectric optical antennas in a wide range of applications such as surface-enhanced Raman spectroscopy \cite{dmitriev2016resonant}, surface-enhanced fluorescence \cite{iwanaga2018all}, directional
light emission \cite{cheng2021superscattering}, nonlinear processes \cite{mobini2021giant}, and Huygens sources \cite{liu2017huygens}. Furthermore, optically induced electric and magnetic Mie resonances make high-index dielectric antennas capable of efficiently engineering the SE rate of quantum emitters with electric or magnetic transition dipole moment \cite{yang2020strong,rolly2012promoting,stamatopoulou2021role,chew1987transition}.
Besides the coexistence of electric and magnetic multipolar resonances at the sub-wavelength scale, unique characteristics such as negligible inherent losses and comparable near-field enhancement with metallic analogs enable high-index dielectric structures to suppress non-radiative
channels and robustify radiative channels in the coupled emitter-dielectric antenna systems. Concentric hollow structures \cite{aslan2021engineering},
Yagi-Uda designs \cite{filonov2012experimental}, dimers \cite{krasnok2016demonstration},
and oligomers \cite{rocco2020giant} have been numerically and experimentally investigated to exhibit the high potential of dielectric sub-wavelength antennas in engineering the SE rate of quantum emitters. 

Among the reported cases exploited to control the SE rate of quantum emitters for specific applications, research has been mostly limited to passive approaches such as design, material, and shape to improve the efficiency of the coupled emitter-photonic systems. Here, we aim to employ active antennas to dynamically control the SE or radiative decay rate of quantum emitters near the proposed antennas. By applying an external static magnetic field, we dynamically tune the scattering response of an optical antenna, leading to the radiative decay rate variations of a quantum emitter. In this respect, we need a material whose optical properties can be modulated upon magnetization \cite{zouros2021three}. Hence, Indium Antimonide (InSb), a III-V semiconductor with unique features of nonreciprocity, magnetoplasmonics, and magnetic tunability in the THz region \cite{chochol2016magneto}, has been exploited to achieve an active sub-wavelength antenna. Without a static magnetic field, the scattering response of
a spherical InSb antenna is divided into plasmonic, epsilon-near-zero
(ENZ), and dielectric regions. By placing an electric dipole (ED) source close to the antenna as the quantum emitter, we observe different types of electromagnetic behaviors
in the coupled emitter-antenna system. Non-radiative channels dominate the plasmonic region due to inherent losses of InSb below the plasma frequency, resulting in the quenching process. A transparency window corresponds to the ENZ region; hence, the quantum emitter does not experience the impact of the antenna. The coexistence of electric and magnetic multipolar resonances is observed in the dielectric region, leading to the radiative decay rate enhancement associated with dominant radiative channels. In the presence of the static magnetic field, the scattering characteristics of the InSb antenna are dramatically changed. The quenching effect becomes much more intense in the plasmonic region. Therefore, the radiative decay rate is negligible compared to the non-radiative one. The ENZ region completely disappears due to the Zeeman-splitting effect of plasmonic modes, and the quantum emitter experiences considerable radiative decay rate variations compared to the unmagnetized case. Despite dynamic control of the emission characteristics of the ED emitter via the magnetized InSb antenna, dominant non-radiative processes degrade the efficiency of the proposed antenna across the studied spectral range. To solve this limitation, as the next step, a hybrid dielectric antenna composed of Silicon (Si) and InSb layers has been proposed to serve as the high-index dielectric antenna, leading to a high radiative decay rate enhancement for the coupled magnetic dipole (MD) and ED emitter-antenna systems. Dual-band SE rate enhancement, robust magnetic response, and high quantum efficiency are important advantages realized by the magnetized hybrid antenna, opening up a possibility for the emergence of tunable magnetic SPSs.

\section{Methodology} 
The spontaneous decay of a two-level system, the so-called quantum emitter, is a radiating process at a quantum level that can be estimated
by Fermi's golden rule \cite{novotny2012principles}. When we consider the interaction of quantum
emitters with optical antennas as open resonators and lossy media,
the normalized total decay rate ($\nicefrac{\varGamma_{\mathrm{tot}}}{\varGamma_{0}}$) must be defined by LDOS in the form
of \cite{novotny2012principles}
\begin{equation}
\frac{\varGamma_{\mathrm{tot}}}{\varGamma_{0}} = \frac{\rho_{n}(\boldsymbol{r}_{0},\omega)}{\rho_{0}(\boldsymbol{r}_{0},\omega)} =
\frac{\boldsymbol{n}_{\boldsymbol{p}}^{\boldsymbol{T}} \;{\mathrm{Im}}\left[\overleftrightarrow{\boldsymbol{G}}_{\!{s}}\left(\boldsymbol{r}_{0},\boldsymbol{r}_{0};\omega\right) + \overleftrightarrow{\boldsymbol{G}}_{\!{0}}\left(\boldsymbol{r}_{0},\boldsymbol{r}_{0};\omega\right)\right] \boldsymbol{n}_{\boldsymbol{p}}}{\boldsymbol{n}_{\boldsymbol{p}}^{\boldsymbol{T}} \; {\mathrm{Im}}\left[\overleftrightarrow{\boldsymbol{G}}_{\!{0}}\left(\boldsymbol{r}_{0},\boldsymbol{r}_{0};\omega\right)\right] \boldsymbol{n}_{\boldsymbol{p}}}
,\label{eq:local density of states}
\end{equation}
where $\rho_{n}\left(\boldsymbol{r}_{0},\omega\right)$ and $\rho_{0}\left(\boldsymbol{r}_{0},\omega\right)$
represent the LDOS at the position $\boldsymbol{r}_{0}$ of the quantum emitter with
a transition frequency $\omega$ in the presence and absence of the
optical sub-wavelength antenna, respectively. The LDOS at the position
of the quantum emitter is computed using the imaginary part of the
Green function \cite{bonod2020controlling}, and $\boldsymbol{n}_{\boldsymbol{p}}$ indicates the polarization's unit vector. Without the antenna,
the LDOS only depends on the electromagnetic field generated by the
quantum emitter $(\overleftrightarrow{\boldsymbol{G}_{0}}\left(\boldsymbol{r}_{0},\boldsymbol{r}_{0};\omega\right))$
at its position. Besides the self-generated field, in the presence of the antenna, the LDOS has another contribution from the electromagnetic field scattered 
by the antenna $(\overleftrightarrow{\boldsymbol{G}_{s}}\left(\boldsymbol{r}_{0},\boldsymbol{r}_{0};\omega\right))$ at the emitter position, showing the impact of the
local field provided by optical sub-wavelength antennas on the total
decay rate modification. The contribution of the LDOS to engineering the decay
rates of a quantum emitter is a common point of classical and quantum
electrodynamics \cite{novotny2012principles}. From a classical point
of view, a quantum emitter can be modeled by a point-like dipole as
an oscillating current source located near an antenna.
Both electrodynamical and classical approaches agree when the intrinsic quantum yield of the point-like quantum source is unity. The total decay rate is defined as the ratio of the
total emitted power of a dipole source in the presence of the antenna
to its total emitted power in the absence of the antenna, in free space, in the form of

\begin{equation}
\frac{\varGamma_{\mathrm{tot}}}{\varGamma_{0}}=\frac{\varGamma_{\mathrm{r}}+\varGamma_{\mathrm{nr}}}{\varGamma_{0}}=\frac{P_{\mathrm{r}}+P_{\mathrm{nr}}}{P_{0}},\label{eq:total decay rate}
\end{equation}
where radiative and non-radiative decay rates are denoted by $\varGamma_{\mathrm{r}}$
and $\varGamma_{\mathrm{nr}}$, respectively. Here, $P_{\mathrm{r}}$ and $P_{\mathrm{nr}}$
are also the scattered and absorbed powers by the antenna, respectively.
The part of the total power incident on the antenna decays into the
non-radiative channel due to the inherent losses of the antenna \cite{novotny2012principles,bonod2020controlling}.
The remaining fraction of the radiation escapes from the coupled emitter-antenna system to free space as far-field radiation. It is thus necessary to calculate the
corresponding powers represented in Eq.~(\ref{eq:total decay rate})
using the Poynting theorem \cite{alhalaby2021enhanced}. The total
emitted power is obtained by integrating the Poynting vector flux
over a surface enclosing the dipole source without the antenna.
The far-field radiation of the coupled emitter-antenna system is computed
by placing a Poynting vector surface integral around the entire system.
The dissipated power is proportional to the power loss density, determined
by a volume integral around the antenna. In this study, we use COMSOL
Multiphysics RF module package 6 to numerically compute the coupled
emitter-antenna system's power rates corresponding to the dipole source's
total decay rate \cite{Comsol,chen2010finite}. Owing to the inherent duality in classical electrodynamics, particularly in Maxwell's equations, the far-field radiation patterns of a quantum emitter possessing a transition magnetic dipole (MD) moment are indistinguishable from those with an electric dipole (ED) moment. Consequently, the same numerical Poynting theorem applicable for EDs can be employed within COMSOL's RF module to compute the total decay rate of an MD emitter. The distinction lies in the necessity to define a point source with a transition MD moment within the simulation model.

\section{Results} 

The primary purpose of this study is to investigate how a static magnetic field applied to an optical antenna dynamically modifies the radiative decay rate of a dipole emitter in the weak-coupling regime. We first consider a spherical InSb antenna embedded in free space to serve as a sub-wavelength THz antenna. The antenna\textquoteright s radius is set to $30$~$\mu\mathrm{m}$, which corresponds to the operation frequency range of 1.2 to 3 THz, matching the range of InSb's plasma frequency.
 The antenna\textquoteright s geometry and frequency range are chosen such that it can support multipolar electromagnetic resonance modes. The dielectric permittivity of InSb becomes anisotropic upon magnetic excitation perpendicular to the incidence plane, enabling the Zeeman-splitting effects on the electronic energy levels of the material proportional to the cyclotron frequency of  $\omega_{c} = \nicefrac{eB}{m^{*}}$ depending on the external magnetic field ($B$), electron charge ($e$) and effective mass of the n-doped InSb ($m^{*}=0.0142m_{0}$).\cite{marquez2020terahertz} (See supporting information for more detail.) 

To gain insight into the light interaction with InSb semiconductor, we evaluate the scattering characteristics of the proposed antenna using multipolar decomposition~\cite{alaee2018electromagnetic}. 
We consider a linearly $x$-polarized plane wave with the wave vector $\mathbf{k}$ propagating along the $y$-direction as illumination. By applying a static magnetic field along the $z$-direction, we induce anisotropy in the dielectric permittivity of InSb, thereby dynamically modifying the scattered-light characteristics (See Fig.~S1 ). 

\begin{figure*}[h!tbp]
\begin{centering}
\includegraphics[width=\linewidth]{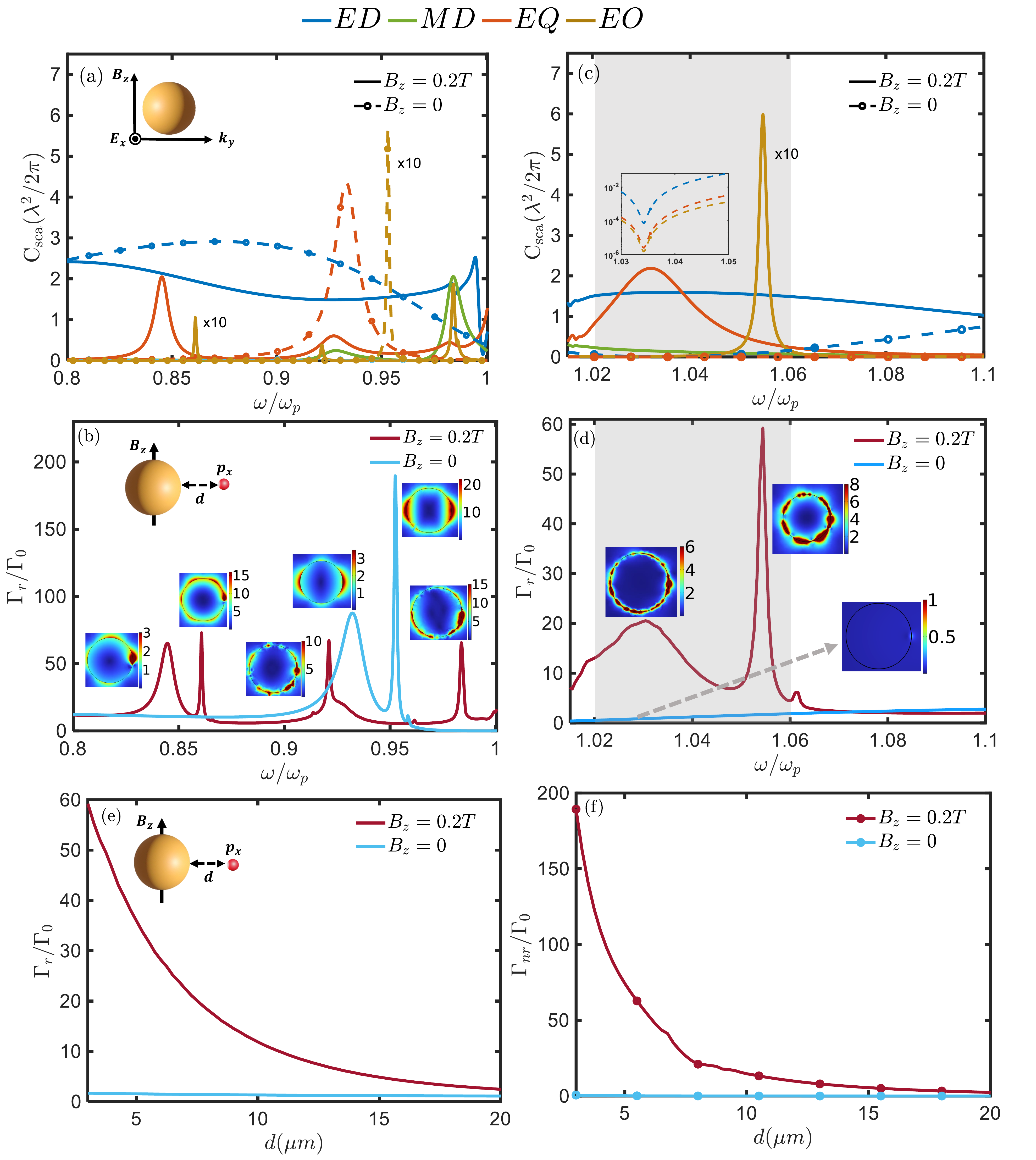}
\par\end{centering}
\caption{(a, b) The comparison of SCS spectra of the spherical InSb antenna
in the presence (solid lines) and absence (dashed lines) of the static magnetic field. The external
magnetic field $(B_{z}=0.2\mathrm{T})$ is applied along the $z$-direction
to dynamically modify the scattering characteristics of the antenna
in the plasmonic (a) and ENZ (b) regions. (c, d) The spectral features
of the radiative decay rate of the $x$-oriented ED source, which is
placed at a distance of $d=3\:\mu\mathrm{m}$ from the antenna. (c) For frequencies
lower than $\omega_{p}$, the coupled system realizes an enhanced
multi-band radiative decay rate upon magnetization, while the unmagnetized
case demonstrates a larger enhancement factor in a single band. (d) Due to the presence of Zeeman-splitting resonances, the radiative
decay rate significantly improves in the ENZ region, highlighted by
the gray area. The insets are the normalized electric-field distributions
around the coupled emitter-antenna system, displaying how the ED source
excites electric resonant modes of the antenna through the radiative
coupling channels. Variations of the radiative (e) and non-radiative (f)
decay rates as a function of coupling distance $(d)$ at the fixed
frequency of $\omega/\omega_{p}=1.05$. \label{fig: figure 1}}
\end{figure*}

As shown in Fig.~1(a), the magnetically induced anisotropy lifts the mode degeneracy, leading to separated plasmonic resonances in the magnetized InSb antenna at frequencies lower than and around $\omega_{p}$. The Zeeman-splitting effect results in new resonant modes whose spectral positions can be predicted \cite{marquez2020terahertz}. These atomic energy level splits are linearly proportional to the Bohr magneton for weak magnetic fields. In the presence of the magnetic field $B_{z}=0.2\mathrm{T}$, a narrow linewidth electric octupole (EO) resonance at $\omega/\omega_{p}=0.951$ splits into multi-band EO resonances, where a peak at $\omega/\omega_{p}=1.055$ considerably amplifies the antenna's scattering response. The magnitudes of the EO resonances have been multiplied by ten to emphasize their critical role in the radiated power, as shown in Fig.~1(a,b). A parent electric quadrupole (EQ) mode at $\omega/\omega_{p}=0.94$ is split into two distinct EQ resonances at $\omega_{-,EQ}/\omega_{p}=0.84$ and $\omega_{+,EQ}/\omega_{p}=1.035$. Moreover, a broad electric dipole (ED) mode is also split into resonances at $\omega_{-, ED}/\omega_{p}=0.8$ and $\omega_{+, ED}/\omega_{p}=,0.99$. Contrary to the unmagnetized scenario, when an applied magnetic field couples with localized surface plasmon resonances, it can excite magneto-plasmonic modes. This results in the contribution of magnetic dipole (MD) resonances to the antenna's spectral response, as depicted in Fig.~1(a,b). The unmagnetized InSb antenna also experiences a plasmonic-to-ENZ transition response at frequencies slightly higher than $\omega_{p}$. This electromagnetic behavior provides invisibility, highlighted by a gray box in Fig.~1(b). The inset depicts the minimum contributions of the multipoles, leading to an almost negligible scattering response, rendering the antenna nearly invisible. In the presence of the magnetic field $B_{z}=0.2\mathrm{T}$, the electromagnetic response of the antenna supports a strong splitting of the EQ mode at $\omega/\omega_{p}=1.035$ and a high $Q$ EO mode at $\omega/\omega_{p}=1.055$, leading to a transition from invisibility to visibility associated with a 60-fold enhancement of the total scattering cross section (SCS). Besides the splitting of the EQ and EO modes, broadband ED and MD resonances also contribute to the SCS of the magnetized InSb antenna.

The contribution of the LDOS to the total decay rate of a quantum emitter implies that local electric field enhancement provided by the plasmonic characteristic of the InSb antenna significantly improves the radiative decay rate of a coupled emitter to the antenna. Such a unique feature motivates us to study the dynamic engineering of a dipole emitter's radiative decay rate in microscale proximity to the spherical InSb antenna beyond a quasi-static approximation. 

Figures~1(c,d) depict the dependence of the radiative decay rate on frequency, where the InSb antenna is fed by an $x$-oriented ED source with the separation distance of $d=3\:\mu\mathrm{m}$. The excited resonant modes of the antenna highly rely on the orientation and nature of the emitter \cite{schmidt2012dielectric}. 
Without the magnetic field, as shown in Fig.~1(c), the radiative decay rate witnesses a significant enhancement in the frequency range from $\nicefrac{\omega}{\omega_{p}}=0.9$ to $0.95$, corresponding to the spectral interval of the parent EQ and EO resonances. The radiative decay rate experiences about 200 times enhancement due to the EO  resonance with a high $Q$ factor, while a broader EQ resonance leads to a lower enhancement. On the other hand, the emitter coupled to the magnetized antenna realizes multi-band radiative decay rate enhancement owing to the contribution of Zeeman-splitting EO, EQ, and ED modes to the radiative coupling channels. The mechanism of the radiative decay rate enhancement around frequencies of $\omega/\omega_{p}=0.85$ and $0.925$ are similar to the unmagnetized case. Lower Zeeman-splitting EO and EQ modes resonantly interact with the transition moment of the ED source, realizing up to 50-fold radiative decay rate enhancement at $\omega/\omega_{p}=0.85$. As shown in Fig.~1(c), another enhancement arises from the EO and EQ modes at $\omega/\omega_{p}=0.925$, where the latter broadens the spectral response of the radiative decay rate. Other radiative decay rate maxima are observed at $\omega/\omega_{p}$= $0.98$, resulting from the higher Zeeman-splitting EO resonance. 

The interesting behavior of the magnetized InSb antenna is observed in Fig.~1(d), where the radiative decay rate enhances up to 60 times due to the resonant interaction of the ED source\textquoteright s transition moment with both the broad EQ and higher Zeeman-splitting EO modes. However, without the magnetic field, the InSb antenna is invisible, and thus the radiative decay rate is equal to one and similar to that in free space across the spectral range highlighted by the gray area.
As mentioned earlier, the antenna can either radiatively or non-radiatively dissipate the emitted energy of the ED source. In the coupled emitter-antenna system, the radiative decay rate channel corresponds to the antenna's scattered electromagnetic field at the emitter position, while the non-radiative decay rate channel is proportional to the absorption losses of the antenna. 

The separation distance between the antenna and the ED source called the coupling distance, is another crucial parameter modifying the total decay rate. Figure 1(e and f) indicates the distance dependence of the radiative and non-radiative decay rates at the fixed frequency of $\omega/\omega_{p}=1.05$ associated
with the transparent window of the InSb antenna. Under zero magnetization, the radiative decay rate is unity, irrespective of the coupling distance,
since the antenna is invisible and the surrounding electromagnetic
environment evokes free space for the dipole source. By applying the
static magnetic field $B_{z}=0.2\:\mathrm{T}$, Zeeman-splitting modes
realize a 60-fold enhancement for the radiative decay rate when
the ED source and the antenna are in close proximity. By increasing
the coupling distance, the resonant interaction of splitting modes
with the ED source\textquoteright s transition moment weakens, and
thus the enhancement factor decreases. For longer distances, the enhancement
factor is unity as the dipole source cannot ``feel'' the antenna (see
Fig.~1(e)). As shown in Fig.~1(f), the unmagnetized coupled emitter-antenna experiences a zero non-radiative
decay rate due to the invisibility of the antenna. Under a biased magnetic field of $B_{z}=0.2\mathrm{T}$, a strong near-field interaction between the ED source's transition moment and the broad ED mode of
the antenna results in the Forster resonant energy transfer (FRET) effect \cite{lackowiczprinciples}, and thus the non-radiative decay rate dramatically enhances for shorter coupling distances. This non-radiative
dipole-dipole interaction becomes ineffective for longer coupling distances, so the non-radiative decay rate falls to zero far from the antenna (see Fig.~1(f)). The magnetized InSb antenna acting as an active photonic device allows us to realize dynamically tunable multi-band SE rate enhancement. However, the total emitted power of the ED source is mainly absorbed by the antenna in the near-field region due to significant Joule heating losses of the InSb material, as shown in Fig.1(f). Dominant non-radiative channels in the coupled system design give rise to a low quantum efficiency for the antenna, impairing the high free-space radiation rate required for SPSs. In the following, we propose an all-dielectric antenna featuring a high quantum efficiency.

Let us consider a hybrid cylindrical antenna composed of silicon (Si) and InSb layers with the same radii of $r=35\:\mu\mathrm{m}$ in the THz regime. The silicone layer, as a high-index and low-loss material, comprises the antenna\textquoteright s upper part with a height of $h_{1}=80\:\mu\mathrm{m}$ to harness the Joule heating losses of the design. The dielectric permittivity of Si is set to $\varepsilon_{\mathrm{Si}}=10.6$ since the material shows a constant permittivity with negligible inherent losses in the studied frequency range. The lower part is made of InSb with a height of $h_{2}=8\:\mu\mathrm{m}$ to supply the magneto-optical properties required for this active antenna. The design principle resides on two aspects. First, the aspect ratio, $\nicefrac{(h_{1}+h_{2})}{r}$, of the cylindrical antenna is optimized to support scattering resonances with high $Q$ factors \cite{rybin2017high}. Second, a given spectral range far from the main absorption band of InSb is chosen such that the material shows moderate positive dielectric permittivity and low losses, as shown in Fig.~S1. When both constituent elements of the antenna exhibit dielectric behaviors, an induced displacement current exceeds the conductive one in the desired frequency range, and thus the all-dielectric antenna can excite robust concurrent electric and magnetic resonances (See Fig.~S3).

\begin{figure*}[h!tbp]
\begin{centering}
\includegraphics[width=\linewidth]{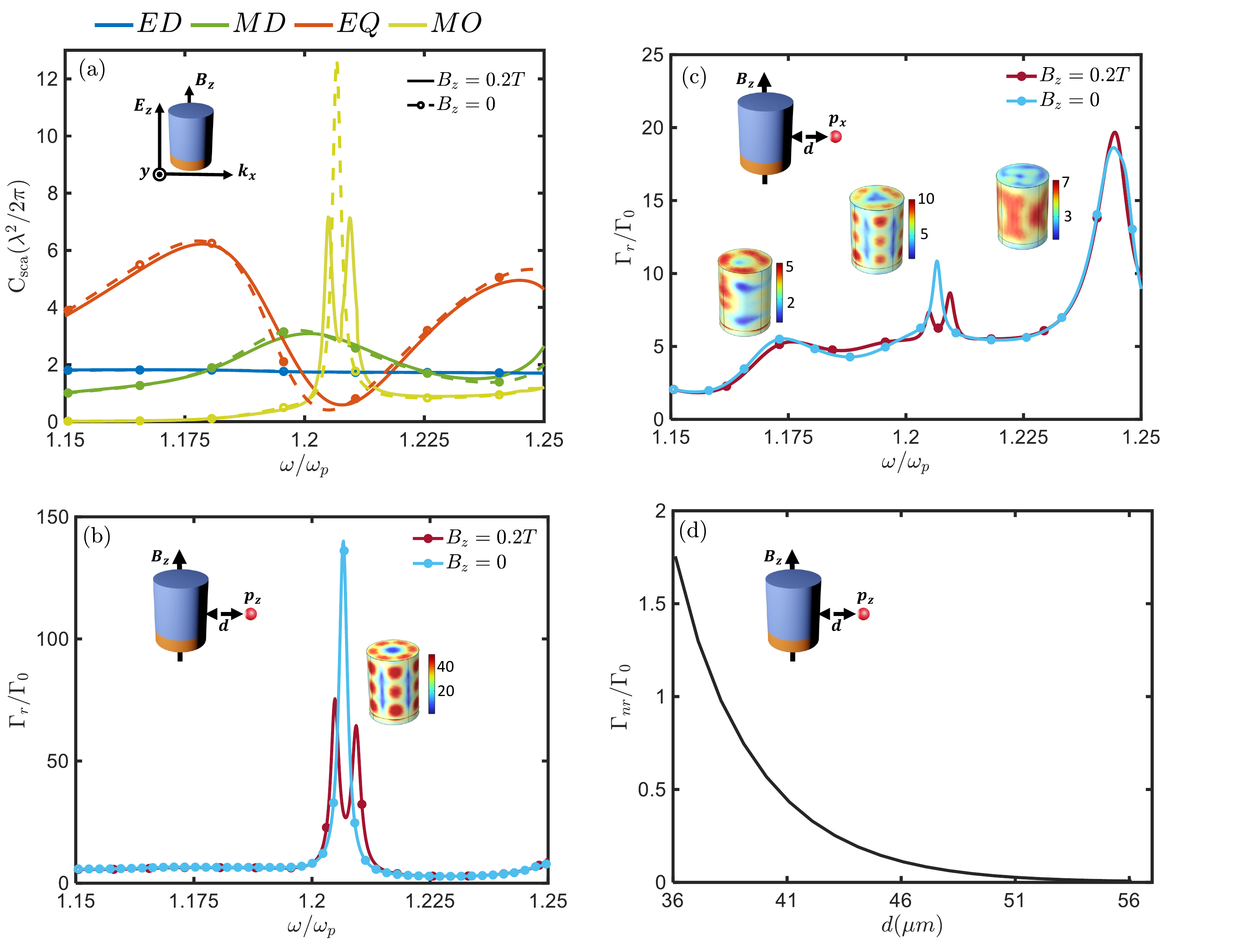}
\par\end{centering}
\caption{(a) The scattering spectra of the hybrid antenna under plane-wave illumination in the presence (solid lines) and absence (dashed lines) of the static magnetic field. By applying the static magnetic field $(B_{z}=0.2\:\mathrm{T})$ along the $z$-direction, multipole moments experience a slight redshift due to the Zeeman effect, while the high-$Q$ parent MO mode splits into two resonant modes in the spectral interval from $\omega/\omega_{p}=1.2$ to $1.22$. (b) The radiative decay rate of the $x$-oriented ED source located near the hybrid antenna with the coupling distance of $d=3\:\mu\mathrm{m}$. The spectral response of the radiative decay rate demonstrates a strong dual-band enhancement factor upon magnetization, whereas the maximum radiative
decay rate enhancement is recorded under zero magnetic bias. The insets display the normalized electric-field distribution of the MO resonance peak excited by the ED emitter. (c) The enhancement of the radiative decay rate for the coupled system is driven by an $z$-oriented ED emitter as a function of frequency. In this orientation, every multipole moment contributes radiatively to the coupling channels. The insets, showing the normalized electric-field distributions for each resonance peak, highlight the relatively weak coupling between the multipole moments and the ED source.
Spectral variations of (d) non-radiative decay rates of the magnetized system upon the $z$-oriented ED emitter excitation at the fixed frequency of $\omega/\omega_{p}=1.205$ corresponding to the resonance frequency of the MO mode. It clearly shows that the non-radiative decay rate of the emitter coupled to the dielectric antenna is significantly lower than that in the coupled emitter-plasmonic antenna system. In this figure, the coupling distance is set to be $3\:\mu\mathrm{m}$.\label{fig: figure5}}
\end{figure*}

In Fig.~\ref{fig: figure5}, we investigate the radiative decay rate variations of an emitter with ED transition, located near the hybrid antenna with the coupling distance of $d=3\:\mu\mathrm{m}$. By applying an external static magnetic field, the antenna's scattering response is modified, enabling one to dynamically control the total decay rate of the coupled ED source. As a comparison, we show the SCS of the magnetized antenna under plane-wave illumination in Fig.~\ref{fig: figure5}(a). In the presence of a magnetic field $B_{z}=0.2\mathrm{T}$, a dominant narrow linewidth magnetic octupole (MO) mode is split into distinct resonances due to the Zeeman effect, while the other broadband multipole resonances experience slight blueshifts. Two MO resonances at frequencies of $\omega_{-,MO}/\omega_{p}=1.202$ and $\omega_{+,MO}/\omega_{p}=1.206$ result from a parent MO resonance at $\omega/\omega_{p}=1.204$. Positioning the $x$-oriented ED source parallel to the hybrid antenna's nearest surface, specifically at the side, enhances the radiative decay rate by a factor of 20 around $\omega/\omega_{p}=1.24$, due to the strong contribution from the broad EQ mode to the radiative coupling channels. In this configuration, the high-$Q$ MO mode has only weak coupling with the ED transition moment of the emitter, resulting in a more modest, 10-fold enhancement of the radiative factor around the resonance frequency of the MO mode (see Fig.~\ref{fig: figure5}(b)). 

Under a biased magnetic field $B_{z}=0.2\:\mathrm{T}$, the coupled system retains multi-band radiative decay rate enhancement, akin to the unmagnetized case. However, the contribution of the MO splitting modes to the radiative coupling channels undergoes further attenuation due to the Zeeman effect. 

Considering the significant impact of the emitter's orientation on decay-rate enhancement, we examine a configuration in which a $z$-oriented ED emitter is aligned parallel to the incident polarization and situated at the side position relative to the antenna, as illustrated in Fig.~2(c). Under zero magnetization, the MO mode strongly contributes to the radiative coupling channels, and the radiative decay rate experiences up to a 140-fold enhancement. As shown in Fig.~\ref{fig: figure5}(c), the coupled system can enhance the dual-band radiative decay rate when the magnetized antenna supports the MO splitting resonances. Insets display the normalized electric field distributions of the coupled system, allowing us to perceive the contribution of multipole moments in the coupling channels. The electric-field profiles verify that the magnetic field application only affects the spectral resonance positions, irrespective of their inherent properties. 

Figure~\ref{fig: figure5}(d) depicts the non-radiative decay rate of the $z$-oriented ED source as a function of the coupling distance ($d$), highlighting that the power absorbed by the antenna at the MO mode's resonance frequency, corresponding to the maximum radiative decay rate, is negligible. This stresses the proficiency of the proposed all-dielectric antenna in suppressing dissipative losses compared to its plasmonic counterpart, as evidenced by the considerably lower non-radiative decay rate relative to the radiative decay rate in the near-field region. Therefore, it is expected to have a high quantum efficiency for the magnetized hybrid antenna that can be exploited for various applications demanding high multi-band radiation.

As demonstrated in Fig.~\ref{fig: figure5}, a magnetic hotspot is achieved due to the MO resonance with a high $Q$ factor, promising highly efficient control of the radiative decay rate of an MD emitter. 
An $x$-oriented MD emitter located perpendicular to the closest surface of the antenna enables the strong radiative coupling channels with the MO resonant mode in the spectral range from $\nicefrac{\omega}{\omega_{p}=}$ $1.2$ to $1.21$. In the coupled MD emitter-antenna system, the radiative decay rate demonstrates a 600-fold enhancement due to a narrow-linewidth MO resonance under zero-bias magnetic field that can be engineered in the presence of the applied magnetic field (See Fig.~3.(a)). Hence, selective emission is another promising advantage of the magnetized hybrid antenna in which spectral bands of the enhanced radiative decay highly depend on the applied magnetic field\textquoteright s strength. Increasing the applied magnetic field makes resonance peaks well-separated across a broader spectral range. The Zeeman-splitting effect is linearly proportional to the cyclotron frequency for magnetic fields $B_{z}<0.3\:\mathrm{T}$, and the split lines are symmetrical with respect to the original resonance peak observed at $B_{z}=0$. For larger magnetic field values, we can not estimate the spectral position of splitting resonances based on Bohr magneton approximation, and new resonance peaks are asymmetrically separated compared to the original resonance. As demonstrated in Fig.3(b), higher-frequency splitting resonances experience considerable blue-shifts while lower ones are slightly red-shifted under a biased magnetic field larger than $B_{z}=0.3\:\mathrm{T}$.

\begin{figure*}[h!tbp]
\begin{centering}
\includegraphics[width=\linewidth]{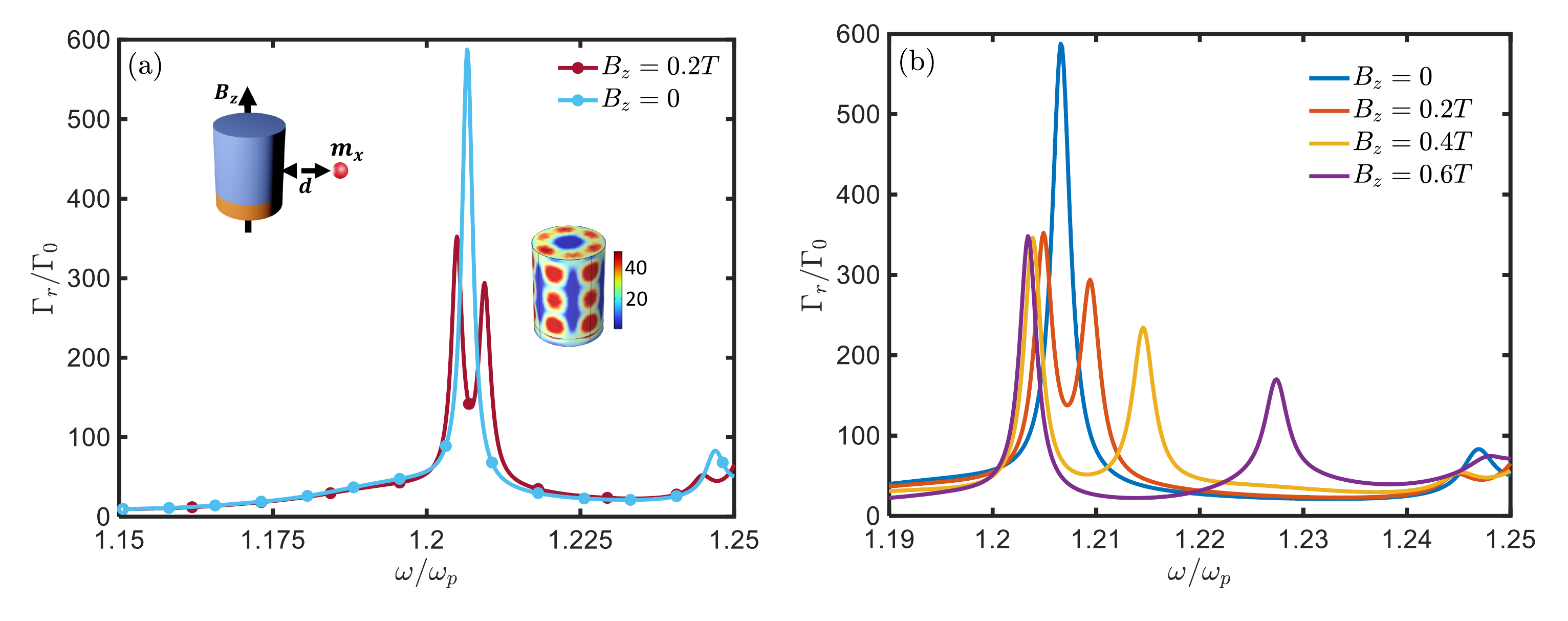}
\par\end{centering}
\caption{(a) Frequency-dependent radiative decay rate of the $x$-oriented MD emitter positioned near the hybrid antenna. The high-$Q$ MO resonance generates substantial radiative coupling with the emitter's transition MD moment, leading to a 600-fold enhancement in the photon production rate. (b). Dynamically tunable dual-band radiative decay rate enhancement of the $x$-oriented MD source employing the applied magnetic field increment in the frequency range corresponding to the high-$Q$ MO mode. \label{fig: figure 1}}
\end{figure*}

As highlighted previously, the LDOS enhancement factor and the normalized radiative and non-radiative decay rates depend on the position, orientation, and type of the dipole source or quantum emitter. Evaluating decay rates based on varied antenna parameters presents complexity and labor intensiveness such that numerical and semi-analytical methods often ignore mutual interactions, leading to suboptimal results. To address this challenge, we employ advanced deep-learning methods recognized for navigating complex and multi-dimensional datasets. We propose a convolutional neural network (CNN) architecture \cite{kiranyaz20211d, ahmadnejad2023tacnet} to discern spatial intricacies within the coupled emitter-antenna system. This CNN design adopts several convolutional layers, employing Rectified Linear Unit (ReLU) activation to determine hierarchical data representations. CNN output undergoes flattening and feeds into dense layers for predictive analysis.

As illustrated in Fig.~4(a), the CNN inputs encompass the spatial coordinates of the quantum emitter, the operational frequency, the value of the applied static magnetic field, and the type of quantum emitter. The initial computational step involves the convolution layer of $C_i = \mathrm{ReLU}(\mathrm{Conv}(X, W_i) + b_i)$, where $C_i$ determines each layer output. Here, $X$ represents the input data, $W_i$ is the weight matrix for the $i^{th}$ filter, and $b_i$ is its corresponding bias. The ReLU function introduces non-linearity, allowing the model to adjust its prediction. This operation is repeated for $i=1$ to $N$, where $N$ is the total number of filters in the convolution layer.

The output is passed through fully connected layers, and the network undergoes a flattening process, converting the 2D data structure from the convolution layer into a 1D vector. This process of transformation is represented as $FC = \mathrm{ReLU}(W_{FC} \times \mathrm{Flatten}(C_N) + b_{FC})$, where $W_{FC}$ represents the weight matrix and $b_{FC}$ is the bias for the fully connected layer. The network concludes with the output layer, which uses the expression of $ \hat{Y} = W_{\text{out}} \times FC + b_{\text{out}} $, where the output $\hat{Y}$ demonstrates the estimated radiative or non-radiative decay rate.

Utilizing our approach, we can determine both the type and optimal positioning of a quantum emitter near an antenna to optimize radiative decay rates. Figure~4(b) illustrates the radiative decay rate for a dipole source positioned variably around the antenna's external surface. Notably, an $x$-oriented MD source placed adjacent to the hybrid antenna at      
($37\:\mu\mathrm{m}, 0\:\mu\mathrm{m}, 40\:\mu\mathrm{m}$) reaches an enhancement factor of 700 at the frequency range corresponding to the MO mode's resonance frequency under zero magnetization. To verify this prediction, Fig.~4(c) displays the simulation result of both radiative and non-radiative decay rates for $m_x$ dipole source at its prime location, mirroring CNN's predictive output. The inset presents the far-field radiation pattern of the maximum value corresponding to the dominant MO mode in a non-magnetized state.

\begin{figure*}[h!tbp]
\begin{centering}
\includegraphics[width=\linewidth]{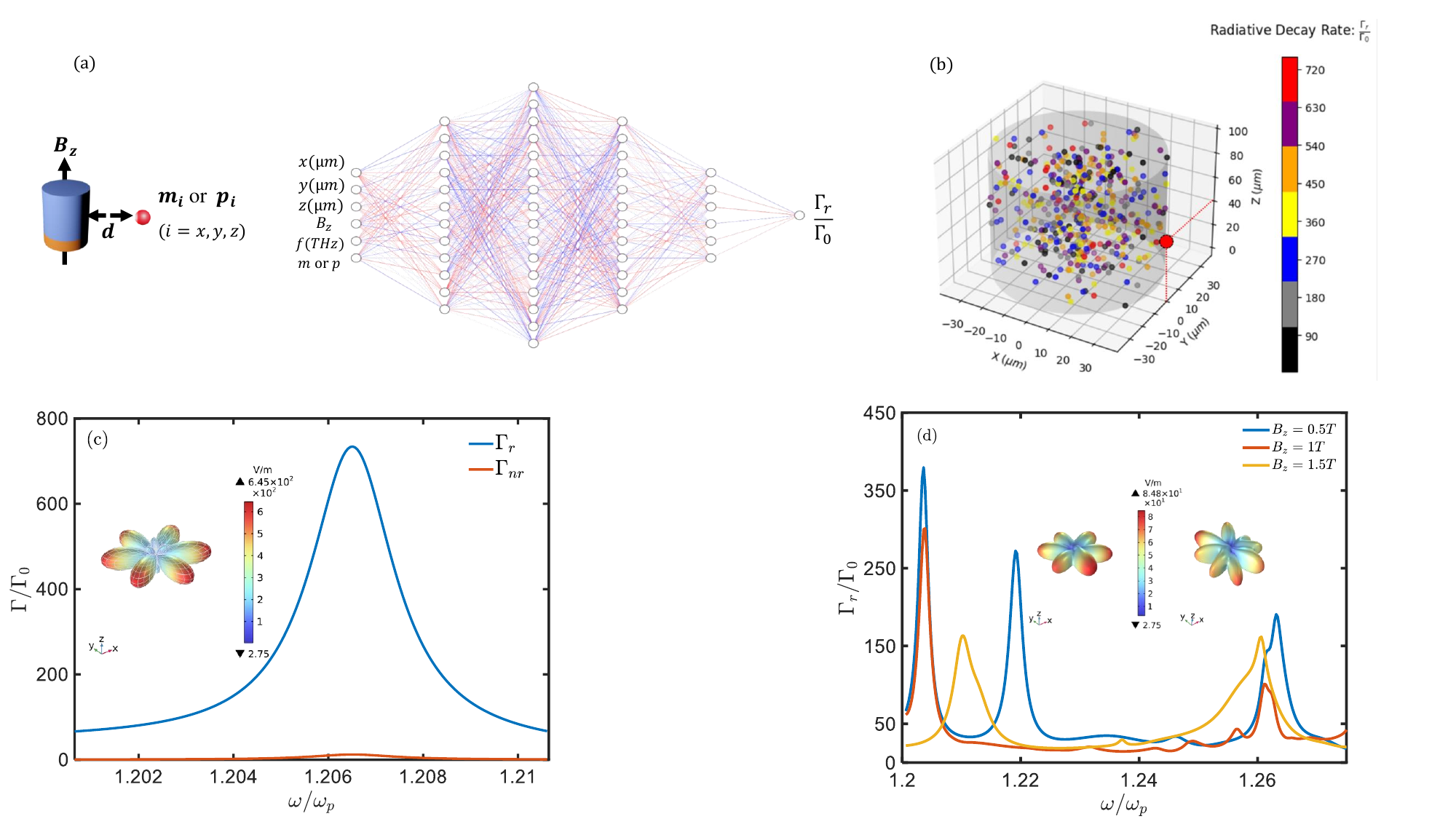}
\par\end{centering}
\caption{ (a) The CNN architecture tailored to identify the ideal position of a variably oriented dipole source that yields the maximum radiative decay rate near the hybrid antenna. This approach holds the constant coupling distance ($d=3\:\mu\mathrm{m}$). (b) The radiative decay rate of the $m_x$ dipole source at varied positions around the hybrid antenna's outer surface. The CNN prediction indicates a seven-fold enhancement factor when $m_x$ is situated at ($37\:\mu\mathrm{m}, 0\:\mu\mathrm{m}, 40\:\mu\mathrm{m}$) within the frequency range corresponding to the MO mode in a non-magnetized state. (c) The radiative and non-radiative decay rates of the $m_x$ dipole source at its optimal position, aligning with the MO mode's resonant frequency in a non-magnetized environment under numerical simulation. The maximum value of the radiative decay rate validates the CNN prediction. The inset highlights the far-field radiation pattern, emphasizing the dominant radiation efficiency of the MO mode into free space. (d) The radiative decay rate variations for three different values of the applied magnetic field. The Zeeman-splitting effect becomes broader with respect to the increment in the applied magnetic field, while the enhancement factor decreases compared to the absence of the magnetic field. Insets exhibit the spatial variations of the far-field electric-field patterns corresponding to radiative decay rate maxima at the resonant frequencies of $\omega/\omega_{p}=1.207, (B_{z}=1.5T)$ and $\omega/\omega_{p}=1.264\: (B_{z}=1 T)$.
\label{fig: figure 4}}
\end{figure*}

\section{Conclusion}

In this study, we numerically investigated the impact of an active antenna on the total decay rate of an emitter. The active antenna enables one to manipulate its scattering response by external agents dynamically. InSb material, a III-V semiconductor with strong magneto-optical
properties in the THz region has been employed to realize the active
antenna that can modify light-matter interaction via an applied static
magnetic field. We used the multipole decomposition method to obtain
the SCS of the antenna under plane-wave illumination. The scattering
response of the antenna allows us to determine which type of emitters
efficiently interact with the antenna. In the first section, we proposed
a spherical InSb antenna to modify the radiative decay rate of an
ED source upon magnetization. We have shown that the radiative decay
rate experiences a multi-band enhancement when the magnetized antenna
has plasmonic features, whereas the maximum enhancement factor of
about 200-fold is observed for the unmagnetized case. Moreover, a
transition from invisibility to visibility has been shown in the ENZ
region such that the transparent antenna turns into a strong scatterer
under a biased magnetic field, and thus the radiative decay rate dramatically
enhances up to 60 times. However, the proposed active antenna suffers
from high Joule heating losses, impairing its quantum efficiency associated
with a low photon production rate. To mitigate this limitation, we
proposed a cylindrical hybrid antenna composed of Si and InSb layers
to act as an all-dielectric active antenna. Due to its robust magnetic
resonance, we have found that the hybrid antenna is very promising
for dynamic control of the radiative decay rate of an MD emitter. Results
show that the magnetized hybrid antenna achieves an enhanced
tunable dual-band radiative decay rate for the MD source, so each
band represents several hundred-fold enhancement factors. Moreover,
we have found that the hybrid antenna harnesses the non-radiative
processes, leading to high quantum efficiency. Such unique features
open up a possibility for tunable SPSs required in quantum information processing. Finally, we employed a modified deep-learning CNN to determine the emitter's position within the coupled system precisely. This position maximizes the radiative decay rate when an $x$-oriented MD source, located at coordinates ($37\:\mu\mathrm{m}, 0\:\mu\mathrm{m}, 40\:\mu\mathrm{m}$), couples with the hybrid antenna. Remarkably, this configuration boosts the radiative decay rate up to 720 times under zero magnetization. This finding was further validated through numerical simulations using Comsol, confirming the prediction of our modified deep-learning network. Offering dynamic splitting bandwidth via the proposed coupled system holds significant potential for designing tunable THz single-photon sources suitable for practical quantum applications.

\newpage
\appendix
\textbf{{\large Supplemental Material}}\\[1em]
\titleformat{\section}[block]{\normalfont\Large\bfseries}{S\arabic{section}\hspace{0.5em}}{0em}{}[]

\section{Dielectric Permittivity of InSb}

We use the Drude model to express the permittivity tensor of InSb at THz frequencies, denoted by

\begin{equation}
\varepsilon\left(\omega\right)=\left[\begin{array}{ccc}
\varepsilon_{xx} & \varepsilon_{xy} & 0\\
\varepsilon_{yx} & \varepsilon_{yy} & 0\\
0 & 0 & \varepsilon_{zz}
\end{array}\right],\label{eq: InSb Permitivity}
\end{equation}
\noindent where the tensor elements of Eq.~(\ref{eq: InSb Permitivity})
are defined as
\begin{equation}
\varepsilon_{zz}\left(\omega\right)=\varepsilon_{\infty}-\frac{\omega_{p}^{2}}{\omega^{2}+i\gamma_{p}\omega}\label{eq: Diagonal ZZ component}
\end{equation}
\begin{equation}
\varepsilon_{xx}\left(\omega\right)=\varepsilon_{yy}\left(\omega\right)=\varepsilon_{\infty}-\frac{\omega_{p}^{2}\left(\omega^{2}+i\gamma_{p}\omega\right)}{\left(\omega^{2}+i\gamma_{p}\omega\right)^{2}-\omega_{c}^{2}\omega^{2}}\label{eq: Diagonal xx and yy component}
\end{equation}
\begin{equation}
\varepsilon_{xy}\left(\omega\right)=-\varepsilon_{yx}\left(\omega\right)=-i\frac{\omega_{p}^{2}\omega_{c}\omega}{\left(\omega^{2}+i\gamma_{p}\omega\right)^{2}-\omega_{c}^{2}\omega^{2}}.\label{eq: off-diagonal component}
\end{equation}

We consider the parameters of an n-doped InSb semiconductor \cite{zouros2020magnetic,zouros2021three}, where $\varepsilon_{\infty}=15.6$ is the background permittivity, $\omega_{p}=12.56\times10^{12}\unitfrac{rad}{s}$
is the plasma frequency, $\gamma_{p}=0.01\omega_{p}$ as the damping constant represents the Ohmic losses of InSb, and $\omega_{c}=\nicefrac{eB_{z}}{m^{*}}$ is the cyclotron frequency depending on the external magnetic field
($B_{z}$), electron charge ($e$) and effective mass ($m^{*}=0.0142m_{0}$).
The low effective mass of InSb results in a strong anisotropy under a weak external magnetic field \cite{chochol2016magneto,zouros2021three,marquez2020terahertz}. Figure~\ref{fig: figure S1} compares the spectral dependency
of InSb\textquoteright s complex permittivities in the presence and absence of the applied magnetic field.  

\renewcommand{\thefigure}{S\arabic{figure}} 
\setcounter{figure}{0}  
\begin{figure*}[h!tbp]
\begin{centering}
\includegraphics[width=0.6\linewidth]{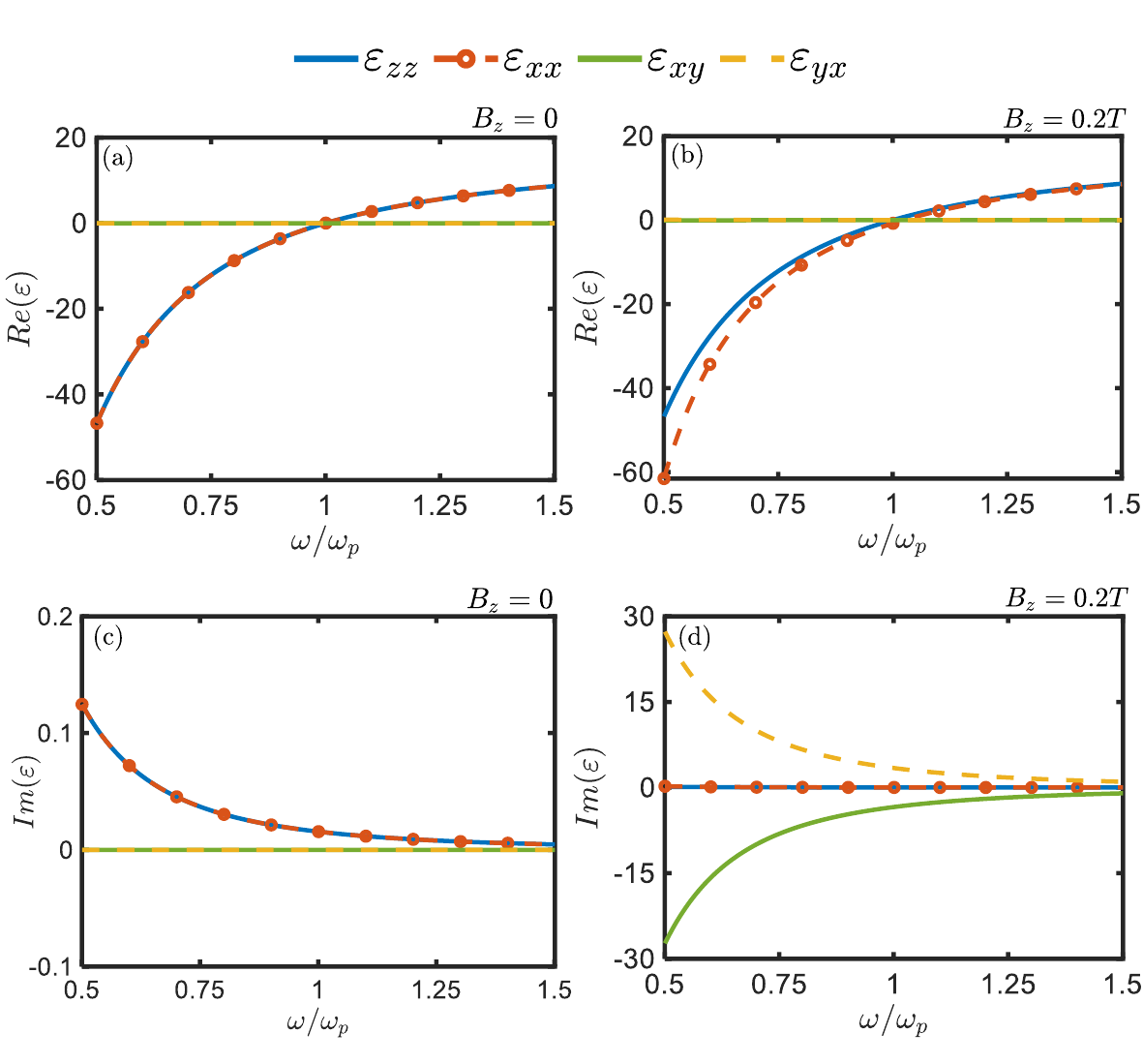}
\par\end{centering}
\caption{(a, c) Real and imaginary parts of InSb's dielectric permittivity versus frequency under zero magnetization $(B_{z}=0)$. Diagonal elements are identical, while off-diagonal ones vanish due to cyclotron frequency dependence. (a) The real part of the diagonal elements is negative at frequencies $\omega/\omega_{p}<1$, representing the plasmonic region of InSb. For frequencies $\omega/\omega_{p}>1$, InSb is
a dielectric material due to the positive real part of its permittivity. (c) The positive imaginary part of diagonal elements is larger in
the plasmonic region, denoting the Joule heating losses of InSb as the plasmonic material. (b, d) Real and imaginary parts of InSb's permittivity as a function of frequency upon magnetization $(B_{z}=0.2\mathrm{T})$.
(b) The real part of diagonal elements increases compared to the unmagnetized
case, and $\varepsilon_{xx}(\omega)$ dissociates from $\varepsilon_{zz}(\omega)$
at lower frequencies. On the other hand, off-diagonal elements remain
almost unchanged, similar to the previous case. (d) In the presence
of the applied magnetic field, the imaginary part of diagonal elements
is negligible, while off-diagonal ones demonstrate considerable variations.
$\varepsilon_{xy}(\omega)$ shows negative values across the studied
spectral range so that it improves the plasmonic behavior of InSb
at frequencies $\omega/\omega_{p}<1$. On the other hand, $\varepsilon_{yx}(\omega)$ is positive throughout the broad spectral range, increasing absorption losses of the material.\label{fig: figure S1} }
\end{figure*}
Without the magnetic field $(B_{z}=0)$, off-diagonal elements of the InSb permittivity disappear due to the cyclotron frequency dependence. On the other hand, diagonal elements are identical, showing negative real parts at lower frequencies than $\omega_{p}$, and positive values for the imaginary parts across the whole spectral range (Fig.~\ref{fig: figure S1}a and c).
Under zero magnetization, the optical response of InSb only depends on the real part of $\varepsilon_{zz}(\omega)$ so that its negative values result in a plasmonic behavior \cite{aghili2022thz}, and the positive ones turn the material into a relatively high-index dielectric \cite{zouros2021three}. Positive values of its imaginary part demonstrate the Ohmic losses of InSb, specifically in the plasmonic region.
In the presence of the magnetic field ($B_{z}=0.2\mathrm{T}$), the real parts of the components remain almost unchanged, and only $\varepsilon_{xx}(\omega)$ increases at lower frequencies compared to the previous case (Fig.~\ref{fig: figure S1}b).
On the other hand, the imaginary parts of the magnetized components experience main variations, in which off-diagonal components demonstrate the same values with different signs. As shown in Fig.~\ref{fig: figure S1}(d), $\varepsilon_{xy}(\omega)$ indicates negative values that can robustify the plasmonic behavior of InSb in agreement with negative real parts of the diagonal components, whereas $\varepsilon_{yx}(\omega)$ increases the Ohmic losses of the material in the plasmonic region due to its considerable positive values.

\section{Scattering Cross Section of the Single Resonant Antenna}

We first need to evaluate the scattering characteristics of the proposed antenna using multipolar decomposition [55]. The incident light can induce an electric conductive or displacement current on the antenna such that the scattered field originates from the induced oscillating current carrying on multipole moments mentioned in Eq.~(11). The respective induced current is described as $J(r,\omega)=i\omega\varepsilon_{0}(\varepsilon(\omega)-1)E(r,\omega)$, where $\varepsilon(\omega)$ is the dielectric permittivity of the antenna and $E(r,\omega)$ describes the electric field distribution around and inside the antenna that can be numerically calculated by the RF module of COMSOL version 6 software. By having the electric-induced current, we can obtain the proposed antenna's total scattering cross-section (SCS) based on the contribution of multipole moments as follows 

\begin{equation}
p_{\alpha}=-\frac{1}{i\omega}\left[\int J_{\alpha}\:dv\:+\frac{k^{2}}{10}\int\left(\mathrm{\mathbf{J.r}}\right)r_{\alpha}-2r^{2}J_{\alpha}\:dv\:+\frac{k^{4}}{280}\int3r^{4}J_{\alpha}-2r^{2}\left(\mathrm{\mathbf{r.J}}\right)r_{\alpha}\:dv\right],\label{eq:dipole moment}
\end{equation}

\begin{equation}
m_{\alpha}=\frac{1}{2}\left[\int\left(\mathrm{\mathbf{r\times J}}\right)_{\alpha}\:dv\:-\frac{k^{2}}{10}\int r^{2}\left(\mathrm{\mathbf{r\times J}}\right)_{\alpha}\:dv\right],\label{eq:magnetic dipole moment}
\end{equation}

\begin{equation}
Q_{\alpha\beta}^{e}=-\frac{1}{i\omega}\left[\int r_{\alpha}J_{\beta}+r_{\beta}J_{\alpha}-\frac{2}{3}\left(\mathrm{\mathbf{r.J}}\right)\delta_{\alpha\beta}\:dv\:+\frac{k^{2}}{42}\int4\left(\mathrm{\mathbf{r.J}}\right)r_{\alpha}r_{\beta}+2\delta\alpha\beta\left(\mathrm{\mathbf{J.r}}\right)r^{2}-5r^{2}\left(r_{\alpha}J_{\beta}+r_{\beta}J_{\alpha}\right)\:dv\right],\label{eq:electric quadropule}
\end{equation}

\begin{equation}
Q_{\alpha\beta}^{m}=\frac{1}{3}\left[\int r_{\alpha}\left(\mathrm{\mathbf{r\times J}}\right)_{\beta}+r_{\beta}\left(\mathrm{\mathbf{r\times J}}\right)_{\alpha}\:dv\:-\frac{k^{2}}{4}\int r^{2}\left(r_{\alpha}\left(\mathrm{\mathbf{r\times J}}\right)_{\beta}+r_{\beta}\left(\mathrm{\mathbf{r\times J}}\right)_{\alpha}\:dv\right)\right],\label{eq:magnetic quadropule}
\end{equation}

\begin{equation}\begin{split}
   O_{\alpha\beta\gamma}^{e}=-\frac{1}{i\omega}\big[\int r_{\alpha}r_{\beta}J_{\gamma}+r_{\beta}r_{\gamma}J_{\alpha}+r_{\gamma}r_{\alpha}J_{\beta}-\frac{1}{5}\delta_{\alpha\beta}\left(r^{2}J_{\gamma}+2\left(\mathrm{\mathbf{r.J}}\right)r_{\gamma}\right)\\
-\frac{1}{5}\delta_{\beta\gamma}\left(r^{2}J_{\alpha}+2\left(\mathrm{\mathbf{r.J}}\right)r_{\alpha}\right)-\frac{1}{5}\delta_{\gamma\alpha}\left(r^{2}J_{\beta}+2\left(\mathrm{\mathbf{r.J}}\right)r_{\beta}\right)\:dv\big],\label{eq:electric octupole}
    \end{split}
\end{equation}

\begin{equation}\begin{split}
O_{\alpha\beta\gamma}^{m}=\frac{1}{24}\big[\int\left(r_{\alpha}r_{\beta}\left(\mathrm{\mathbf{r\times J}}\right)_{\gamma}+r_{\beta}r_{\gamma}\left(\mathrm{\mathbf{r\times J}}\right)_{\alpha}+r_{\gamma}r_{\alpha}\left(\mathrm{\mathbf{r\times J}}\right)_{\beta}\right)\\
-\frac{1}{5}\delta_{\alpha\beta}r^{2}\left(\mathrm{\mathbf{r\times J}}\right)_{\gamma}-\frac{1}{5}\delta_{\beta\gamma}r^{2}\left(\mathrm{\mathbf{r\times J}}\right)_{\alpha}-\frac{1}{5}\delta_{\gamma\alpha}r^{2}\left(\mathrm{\mathbf{r\times J}}\right)_{\beta}\:dv\big],\label{eq:magnetic octupole}
\end{split}
\end{equation}
and the scattering cross-section from the antenna can be expressed in terms of multipole moments provided by Eqns.~(5) to (10) as
\begin{equation}\begin{split}
 C_{\mathrm{sca}}^{\,\mathrm{tot}}=\frac{k^{4}}{6\pi\varepsilon_{0}^{2}\left|E_{\mathrm{inc}}\right|^{2}}\left|p_{\alpha}\right|^{2}+\frac{k^{4}}{6\pi c^{2}\varepsilon_{0}^{2}\left|E_{\mathrm{inc}}\right|^{2}}\left|m_{\alpha}\right|^{2}+\frac{k^{6}}{160\pi\varepsilon_{0}^{2}\left|E_{\mathrm{inc}}\right|^{2}}\sum_{\alpha\beta}\left|Q_{\alpha\beta}^{e}\right|^{2}\\
+\frac{k^{6}}{80\pi c^{2}\varepsilon_{0}^{2}\left|E_{\mathrm{inc}}\right|^{2}}\sum_{\alpha\beta}\left|Q_{\alpha\beta}^{m}\right|^{2}+\frac{k^{8}}{1890\pi\varepsilon_{0}^{2}\left|E_{\mathrm{inc}}\right|^{2}}\sum_{\alpha\beta\gamma}\left|O_{\alpha\beta\gamma}^{e}\right|^{2}+\frac{k^{8}}{1890\pi c^{2}\varepsilon_{0}^{2}\left|E_{\mathrm{inc}}\right|^{2}}\sum_{\alpha\beta\gamma}\left|O_{\alpha\beta\gamma}^{m}\right|^{2}.\label{eq:total scattering}
\end{split}
\end{equation}

Here $p_{\alpha}$/$m_{\alpha}$, $Q_{\alpha\beta}^{e}$/$Q_{\alpha\beta}^{m}$, and $O_{\alpha\beta\gamma}^{e}$/$O_{\alpha\beta\gamma}^{m}$ are electric/magnetic dipole moments, electric/magnetic quadrupole moments, and electric/magnetic octupole moments, respectively. $c$ is the speed of light, $E_{0}$ represents the electric field magnitude of the incident light, and $\alpha,\beta,\gamma=x,y,z$.
\subsection{Spherical InSb Antenna}

We consider a linearly $x$-polarized plane wave propagating with the wavevector $k$ along the $y$-direction as illumination. By applying a static magnetic field, we dynamically modify the scattered light characteristics of the antenna owing to the anisotropic behavior observed in InSb dielectric permittivity (See Fig.~1).
\renewcommand{\thefigure}{S\arabic{figure}} 
\setcounter{figure}{1}  
\begin{figure*}[h!tbp]
\begin{centering}
\includegraphics[width=1\linewidth]{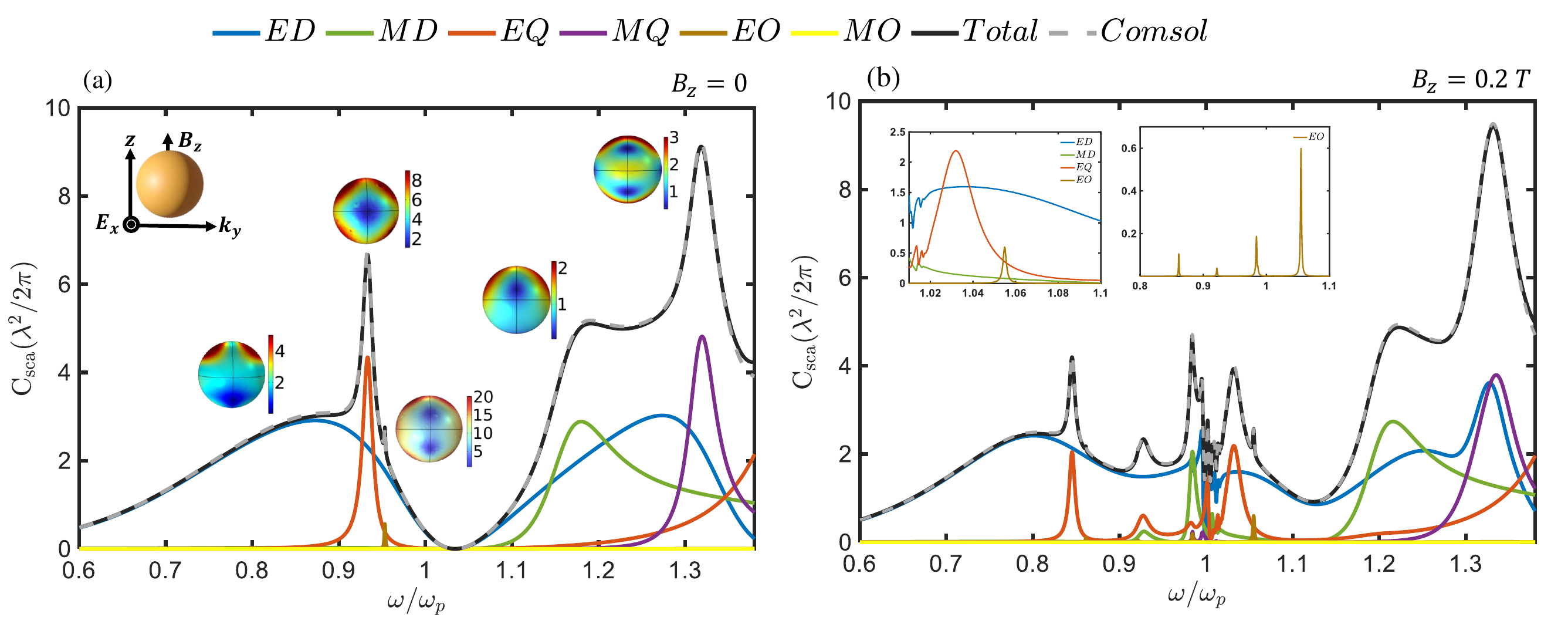}
\par\end{centering}
\caption{The scattering characteristics of the spherical InSb antenna with the radius $r=30\mu\mathrm{m}$
under $x$-polarized plane wave illumination with the wavevector $k_{y}$, (a) in the absence and (b) presence of the applied magnetic field. (a) The total SCS of the antenna indicates five resonance peaks arising from
the contribution of dominant ED, MD, EQ, MQ, and EO modes. The insets demonstrate the antenna's instantaneous normalized electric field distributions corresponding to the total SCS maxima, further clarifying the nature
of electromagnetic resonant modes. (b) The scattering characteristics of the antenna are modified upon magnetization where the Zeeman splitting effect makes visibility from Invisibility around $\nicefrac{\omega}{\omega_{p}}=1$. The insets specifically focus on the SCS of the magnetized antenna in the ENZ region. Here, the high-$Q$ multiband EO resonant modes significantly influence the antenna's scattering response, thereby enhancing light manipulation. \label{fig: figure S3} }
\end{figure*}
\newpage
Figure~\ref{fig: figure S3} shows the SCS of the InSb antenna, allowing us to identify the electromagnetic
resonant modes excited in the system. The spectral response of the SCS
is divided into three specific regions corresponding to InSb\textquoteright s
dielectric permittivity shown in Fig.~\ref{fig: figure S1}(a). The
induced electric conductive current excites a broad ED, a narrow EQ, and an ultra-narrow linewidth EO with a very small magnitude corresponding to negative values of InSb\textquoteright permittivity to realize
a plasmonic antenna in the studied frequency range of $\nicefrac{\omega}{\omega_{p}}<1$.
The total SCS of the antenna at frequencies slightly higher than $\omega_{p}$
nearly vanishes to provide a transparent region associated with near-zero
values of InSb permittivity. An induced circular displacement current
is excited inside the antenna at higher frequencies, so the total SCS
also has contributions from magnetic dipole (MD) and magnetic quadrupole
(MQ) modes. The concurrent support of multipolar electric and magnetic
resonances realizes a dielectric antenna associated with positive
real parts of InSb permittivity. The scattering response of the antenna is dramatically enhanced in the transparent region under a biased magnetic field due to the contribution of the splitting EQ and EO resonant modes, as shown in Fig.~\ref{fig: figure S3}(b).

\subsection{ Hybrid Antenna}

In this section, we scrutinize the scattering response of the hybrid antenna excited by a linearly $z$-polarized plane wave impinging along the $x$-axis, as shown in Fig.~\ref{fig: figure 4}. The antenna is made of a lower InSb layer and an upper Si layer with a radius of $35\mu\mathrm{m}$ and a height of $8\mu\mathrm{m}$ and $80\mu\mathrm{m}$ respectively. Under zero magnetization, the multipolar decomposition of SCS reveals a resonance with a high $Q$ factor in the spectral interval around $\omega/\omega_{p}=1.22$, corresponding to contributions of the dominant MO mode, which is splitted into two resonant modes with high-$Q$ factors upon weak magnetization. Insets demonstrate the normalized instantaneous electric field distributions of the total SCS maxima, allowing us
to elucidate how the incident wave resonantly interacts with the antenna. The anti-parallel electric field\textquoteright polarizability at opposite sides of the dielectric cylinder induces a circular displacement current, leading to a narrow linewidth MO resonance inside the particle. 
\renewcommand{\thefigure}{S\arabic{figure}} 
\setcounter{figure}{2}  
\begin{figure*}[h!tbp]
\begin{centering}
\includegraphics[width=1\linewidth]{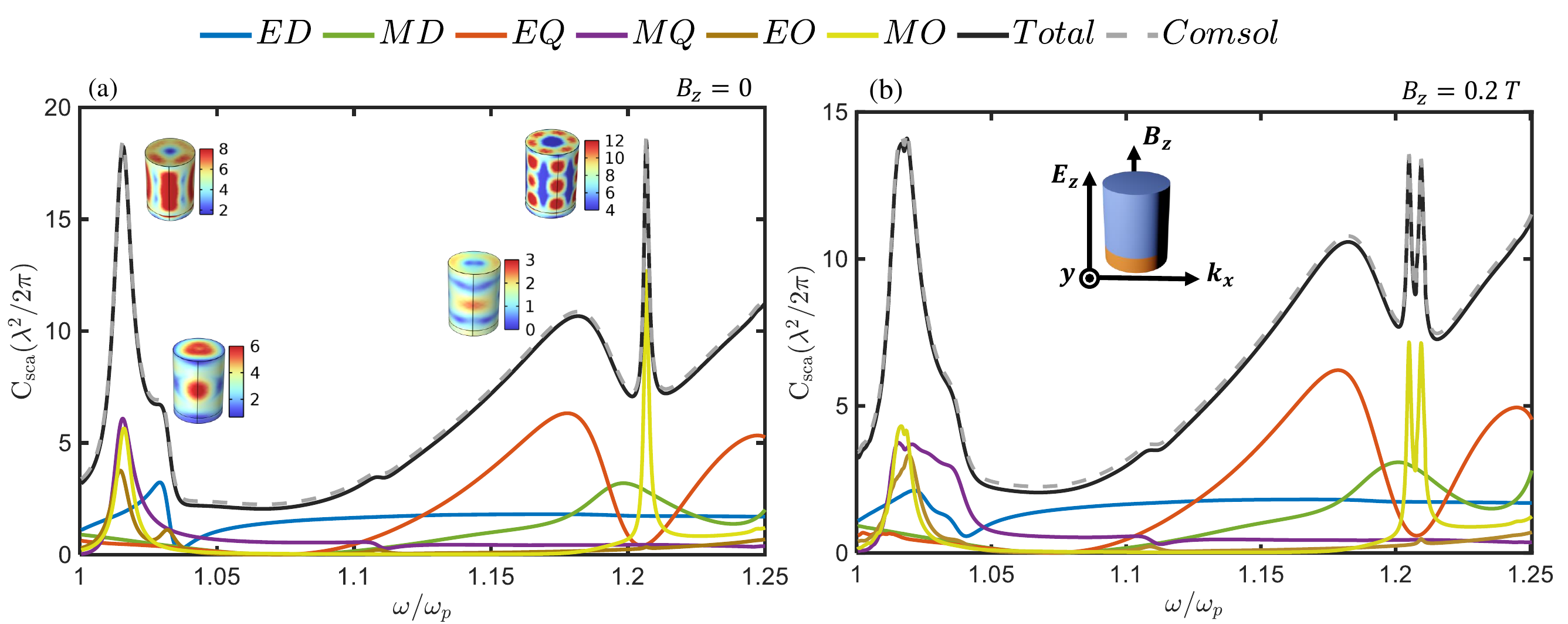}
\par\end{centering}
\caption{The SCS of the hybrid cylindrical antenna under $z$-polarized plane wave illumination with the wavevector $k_{x}$ (a) in the absence and (b) presence of the applied static magnetic field. The scattering response of the antenna supports all multipole moments across the studied spectral range in which the contribution of the MO mode is significant and experiences the Zeeman splitting effect upon magnetization, leading to the splitted MO resonant modes, as shown in panel (b). The upper part of the antenna is made of a Si layer with the height and radius of $80$ and $35\mu\mathrm{m}$, respectively. InSb layer comprises the lower part of the antenna with the height and radius of $8$ and $35\mu\mathrm{m}$,
respectively. Insets show the instantaneous normalized electric field
distributions around the antenna corresponding to the total SCS maxima.
\label{fig: figure 4}}
\end{figure*}

\section{Optimization Method Algorithm}

Our methodology employs stochastic gradient descent (SGD) to refine our model's parameters iteratively. SGD fine-tunes the CNN's weights and biases by minimizing the MSE loss. Given the intricate high-dimensional nature of the coupled system features, we use a genetic algorithm to pinpoint the most relevant feature subset. This algorithm evolves a population of feature sets over multiple iterations, leveraging crossover and mutation techniques. The effectiveness of each feature set is assessed through the accuracy of CNN's predictions.

In our validation, we conduct rigorous tests on a contemporary dataset encompassing diverse antenna systems. Our deep learning model outperforms conventional regression techniques. Furthermore, the genetic algorithm adaptly discerns the paramount features crucial for precise decay rate prediction. Overall, integrating deep learning with genetic algorithm-driven feature selection presents a significant advancement in addressing challenges in coupled emitter-antenna systems (See. Algorithm 1).

\begin{algorithm}[H]
\caption{Enhanced Deep Learning and Genetic Algorithms for Radiative Decay Rate Prediction}
\begin{algorithmic}[1]
\State \textbf{Data Preprocessing:}
\State Normalize spatial coordinates ($x$, $y$, $z$)
\State Apply one-hot encoding for frequency and applied magnetic field
\State \textbf{Neural Network Initialization:}
\State Set up neural network architecture with random weights and biases
\State \textbf{Genetic Algorithm Parameters:}
\State Set population size $P$, number of generations $G$, and mutation rate $M$
\State Initialize population of feature sets $F$ with $P$ individuals
\State \textbf{Deep Learning Training:}
\State Define the number of epochs and learning rate
\For {each epoch}
    \State Compute predicted radiative decay rate $\hat{Y}$
    \State Compute MSE loss: $MSE = \frac{1}{n} \sum_{i=1}^{n} (Y_i - \hat{Y}_i)^2$
    \State Update model parameters via optimizer
\EndFor
\State \textbf{Genetic Optimization:}
\For{generation $g = 1$ to $G$}
    \State Evaluate fitness of each feature set $f \in F$ using the neural network
    \State Select best-performing individuals for mating pool $MatingPool$
    \State Populate $NewF$ by applying genetic operations on $MatingPool$
    \For {each feature set $f \in NewF$}
        \If{random value $<$ $M$}
            \State Introduce mutation to $f$
        \EndIf
    \EndFor
    \State Update population: $F \leftarrow NewF$
\EndFor
\State \textbf{Optimal Features:}
\State Select feature set with the topmost fitness value from the final population
\State \textbf{Final Prediction:}
\State Predict radiative decay rate using the selected feature set
\end{algorithmic}
\end{algorithm}

\bibliographystyle{unsrt}
\bibliography{sample}

\end{document}